\documentclass[twocolumn,showpacs,preprintnumbers]{revtex4}
\usepackage{dcolumn}
\usepackage{bm}
\usepackage{amsmath}
\usepackage{hyperref}
\usepackage{graphicx}
\usepackage{subfigure}

\newcommand{\td}{\mathrm{d}}
\newcommand{\te}{\mathrm{e}}
\newcommand{\ti}{\mathrm{i}}
\newcommand{\ua}{\uparrow}
\newcommand{\da}{\downarrow}

\begin{document}

\title{Theory of magnon--skyrmion scattering in chiral magnets}
\author{Junichi Iwasaki$^1$}
\email{iwasaki@appi.t.u-tokyo.ac.jp}
\author{Aron J. Beekman$^2$}
\author{Naoto Nagaosa$^{1,2}$}
\email{nagaosa@ap.t.u-tokyo.ac.jp}
\affiliation{$^1$ Department of Applied Physics, The University of Tokyo, 7-3-1, 
Hongo, Bunkyo-ku, Tokyo 113-8656, Japan}
\affiliation{$^2$RIKEN Center for Emergent Matter Science (CEMS), Wako, Saitama 351-0198, Japan}

\begin{abstract}
We study theoretically the dynamics of magnons in the presence of a single skyrmion in 
chiral magnets featuring Dzyaloshinskii--Moriya interaction. We show by micromagnetic simulations that the scattering 
process of magnons by a skyrmion can be clearly defined although both originate in the 
common spins. We find that (i) the magnons are deflected by a skyrmion, with the angle 
strongly dependent on the magnon wavenumber due to the effective magnetic field of the topological texture,
and (ii) the skyrmion motion is driven by 
magnon scattering through exchange of the momenta between the magnons and a skyrmion:
the total momentum is conserved. This demonstrates that the skyrmion has a 
well-defined, though highly non-Newtonian, momentum.
\end{abstract}

\pacs{73.43.Cd,72.25.-b,72.80.-r}
\maketitle

{\em Introduction. }
The particle in field theory is not a trivial concept. There are infinite possible field 
configurations, some of which are identified as {\em particles}.  Starting from the vacuum 
state, which often corresponds
to a trivial field configuration such as the perfect alignment of spins, the small perturbative 
deviations are naturally described by noninteracting plane waves. Localized particles, on the 
other hand, require nonlinear interaction and non-perturbative effects. In magnets, the 
spin waves or magnons correspond to the former, and localized spin textures to the latter.
However, the separation into waves and particles cannot always be clearly distinguished, 
and the particles can be dissociated into waves when they collide with each other.    
Here, the topology plays an essential role: it gives the identity and stability of 
spin textures such as domain walls and skyrmions and also determines the scattering 
process with the magnons as will be shown below.

The skyrmion was first proposed as a model for hadrons in nuclear physics~\cite{Skyrme61, Skyrme62} 
and has been discovered in a variety of condensed matter systems~\cite{Wright89, Sondhi93, Ho98}, 
most recently in magnets with Dzyaloshinskii--Moriya (DM) interaction~\cite{Mulbauer09,Yu10,Heinze11,Seki12}. 
Here it is a topological spin texture characterized by the skyrmion number $Q$.
The skyrmion has very long lifetime because of topological protection, 
i.e. any continuous deformation of the field configuration cannot change the 
skyrmion number. On the other hand, the low-energy excitations in magnets are magnons: propagating 
small disturbances in the underlying spin texture. Then, a natural question to ask is how magnons 
interact with skyrmions. It has been known that the motion of a domain wall in ferromagnets 
can be induced by magnons: the domain wall moves against the direction of the magnon 
current~\cite{Yan11,Hinzke11,Kovalev12}. Recently the skyrmion version of the magnon-induced motion 
has been studied~\cite{Kong13}, when magnons are produced by a temperature gradient. 
However, the elementary process involving a single skyrmion and magnons has not been 
studied up to now. The only work on magnon-skyrmion dynamics we are aware 
of (ref.~\cite{Ferrer00}) precludes from the outset, in the context of quantum Hall systems, 
any skew-scattering, which does not agree with the observations in chiral magnets. 
Another work considered magnon scattering off skyrmions in time-reversal invariant 
systems \cite{Walliser99}.

The skyrmion is characterized by a spin gauge field ${\bf a}$
and carries an emergent magnetic flux ${\bf b} = \nabla \times {\bf a}$
associated with the solid angle subtended by the spins. This spin gauge field
${\bf a}$ is coupled to the conduction electrons, which results in nontrivial effects
such as the spin transfer torque driven skyrmion motion and topological Hall effect.  
Surprisingly, a tiny current density $\sim 10^6$ A/m$^2$ can drive the 
motion of skyrmion crystal via spin transfer torque~\cite{Jonietz10,Yu12}, 
which is orders of magnitude smaller than that in domain wall motion in ferromagnets
($10^{10}$--$10^{12}$ A/m$^2$)~\cite{Parkin08,Yamanouchi04}.
This has been attributed to the Magnus force acting on the skyrmion and its flexible shape 
deformation reducing the threshold current~\cite{Schulz12,Iwasaki13}.
An interesting recent development is the discovery of skyrmions in an insulating 
magnet Cu$_2$OSeO$_3$~\cite{Seki12,White12,Liu13}, where the electric-field-induced 
motion is associated with multiferroic behaviour.
It is expected that in this insulating system, the only low-energy relevant excitations
are the magnons, and the interaction between magnon and skyrmion becomes especially relevant.

\begin{figure*}[t]
\begin{center}
\includegraphics[width=0.95\hsize]{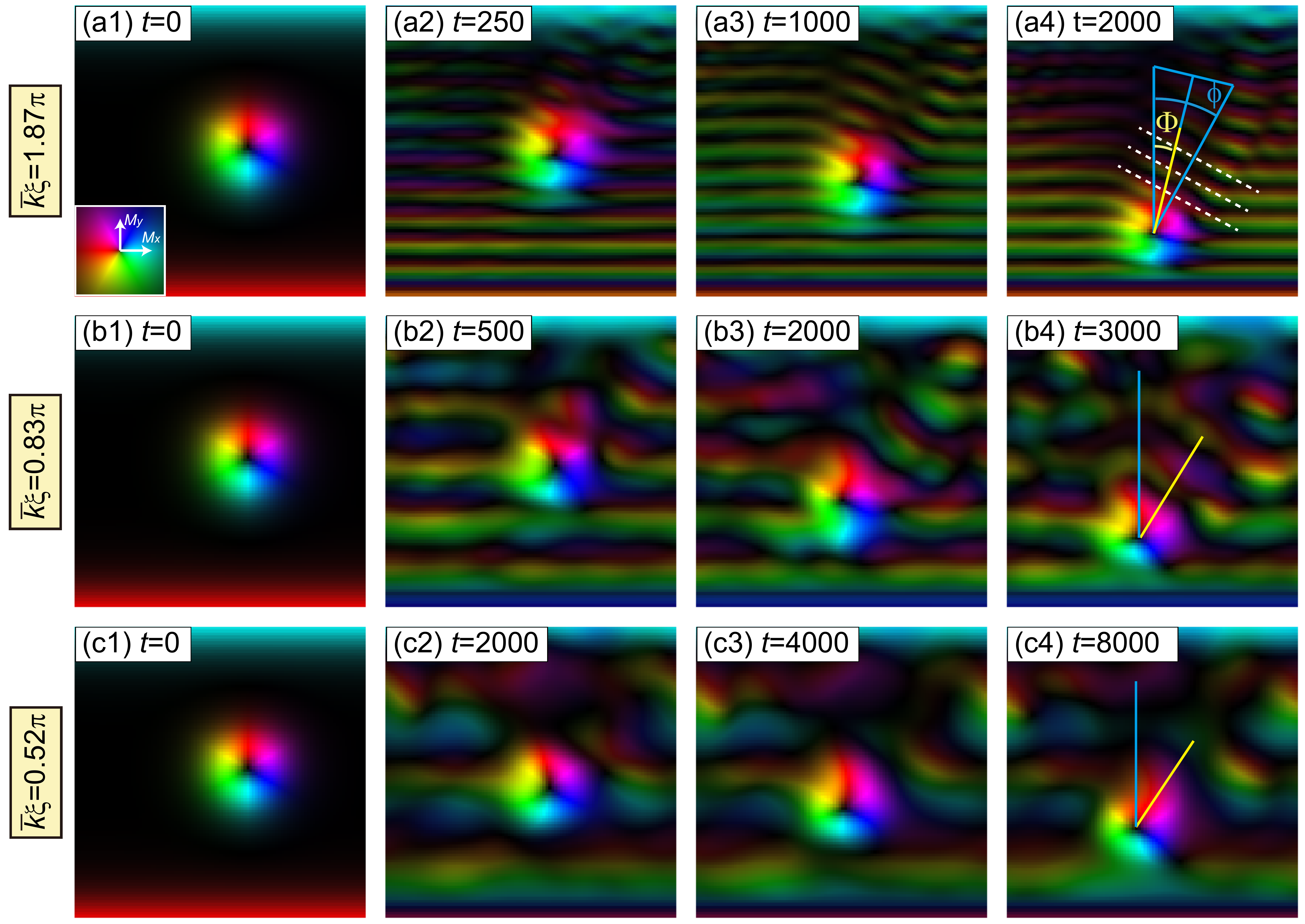}
\end{center}
\caption{
{  Snapshots of scattering processes with three different wavenumbers.}
(a1)--(a4): $\bar{k}\xi \simeq 1.87\pi$, 
(b1)--(b4): $\bar{k}\xi \simeq 0.83\pi$, 
(c1)--(c4): $\bar{k}\xi \simeq 0.52\pi$, timesteps of the snapshots as indicated. 
The inset of (a1) shows the colour representation of the in-plane spin component in 
$(xy)$ spin space. In (a4), (b4) and (c4), the vertical blue line denotes the incoming magnon 
direction. For the higher wavenumbers we can clearly identify the skew scattering of the 
magnons. In (a4) the white dashed lines indicate the equal phase contour of 
the scattered magnons, and blue line perpendicular to those defines the scattering skew angle $\bar{\varphi}$. 
The yellow lines represent the path traversed by the skyrmion also clearly showing skew 
scattering over an angle $\Phi$. Hence we see that the skyrmion skew angle is nearly 
half of the magnon skew angle as expected from the conservation of the momentum. 
}
\label{fig:snapshots}
\end{figure*}

In this paper, we study the scattering process of a magnon by a skyrmion by solving 
numerically the Landau--Lifshitz--Gilbert (LLG) equation for magnons with the center of 
wavenumbers $k$ incident on a 
skyrmion of size $\xi$. The simulations clearly show wavenumber-dependent skew-scattering of 
the magnon, and furthermore similar large Hall angle skyrmion motion due to the back action.
This process is well analyzed in terms of the momentum conservation, strongly indicating that the skyrmion is particle-like with a well-defined momentum and low mass.
By mapping the situation to a charged particle scattered by a tube of magnetic flux we show 
that the principal contribution to skew-scattering is the emergent Lorentz force generated by 
the skyrmion.

\begin{figure*}[t]
\begin{center}
\subfigure[]{\includegraphics[height=0.24\hsize]{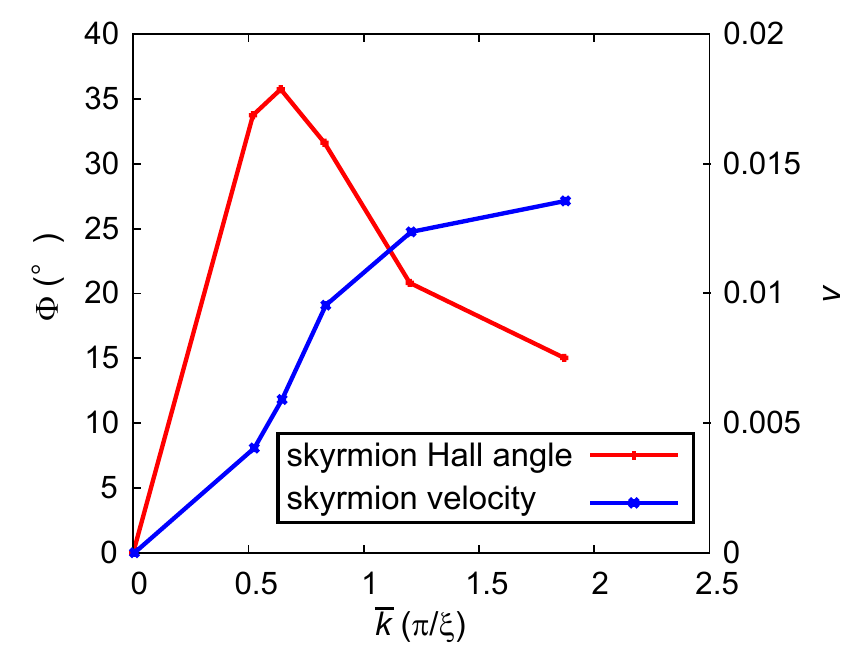}}
\subfigure[]{\includegraphics[height=0.24\hsize]{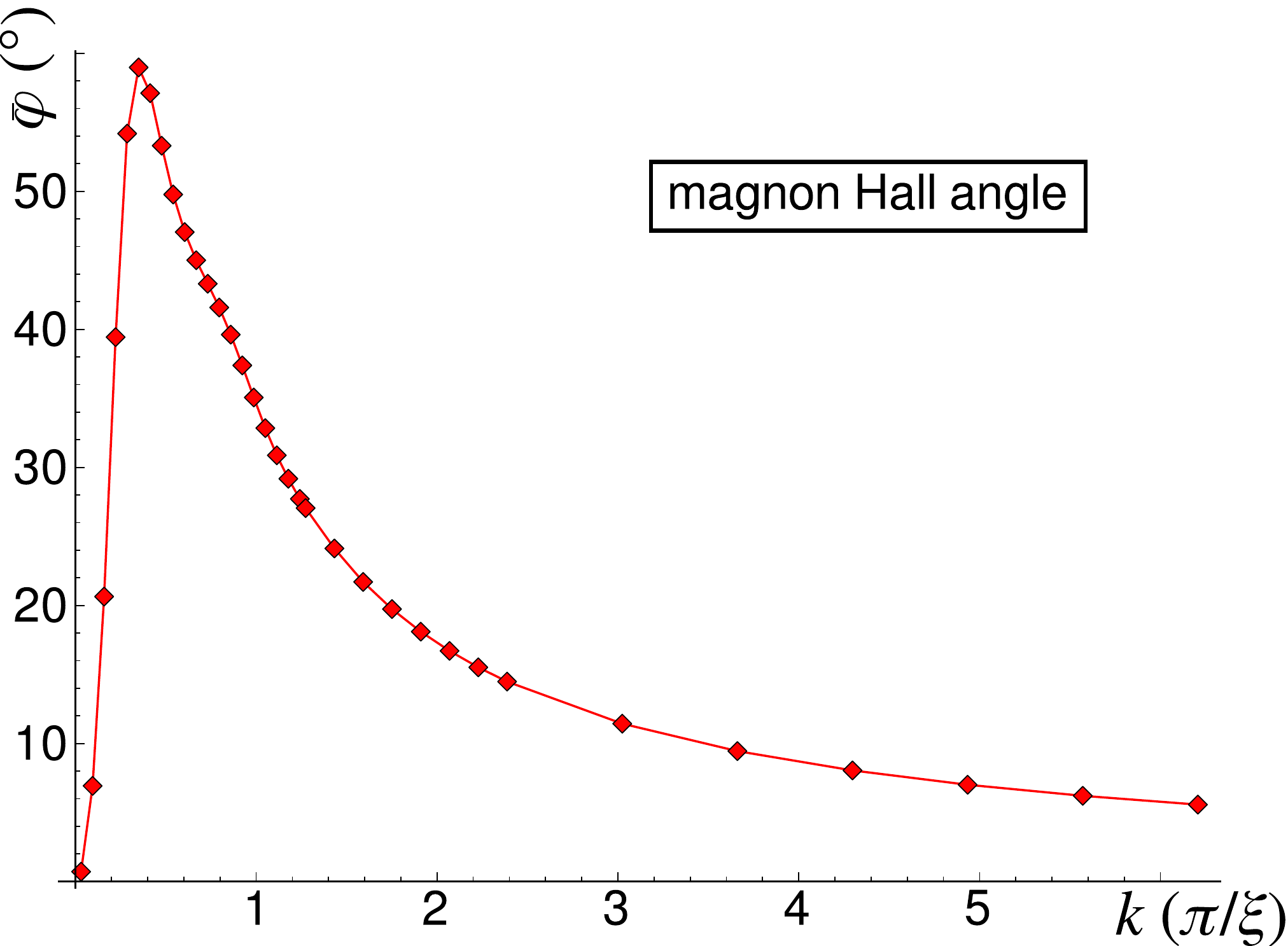}}
\subfigure[]{\includegraphics[height=0.24\hsize]{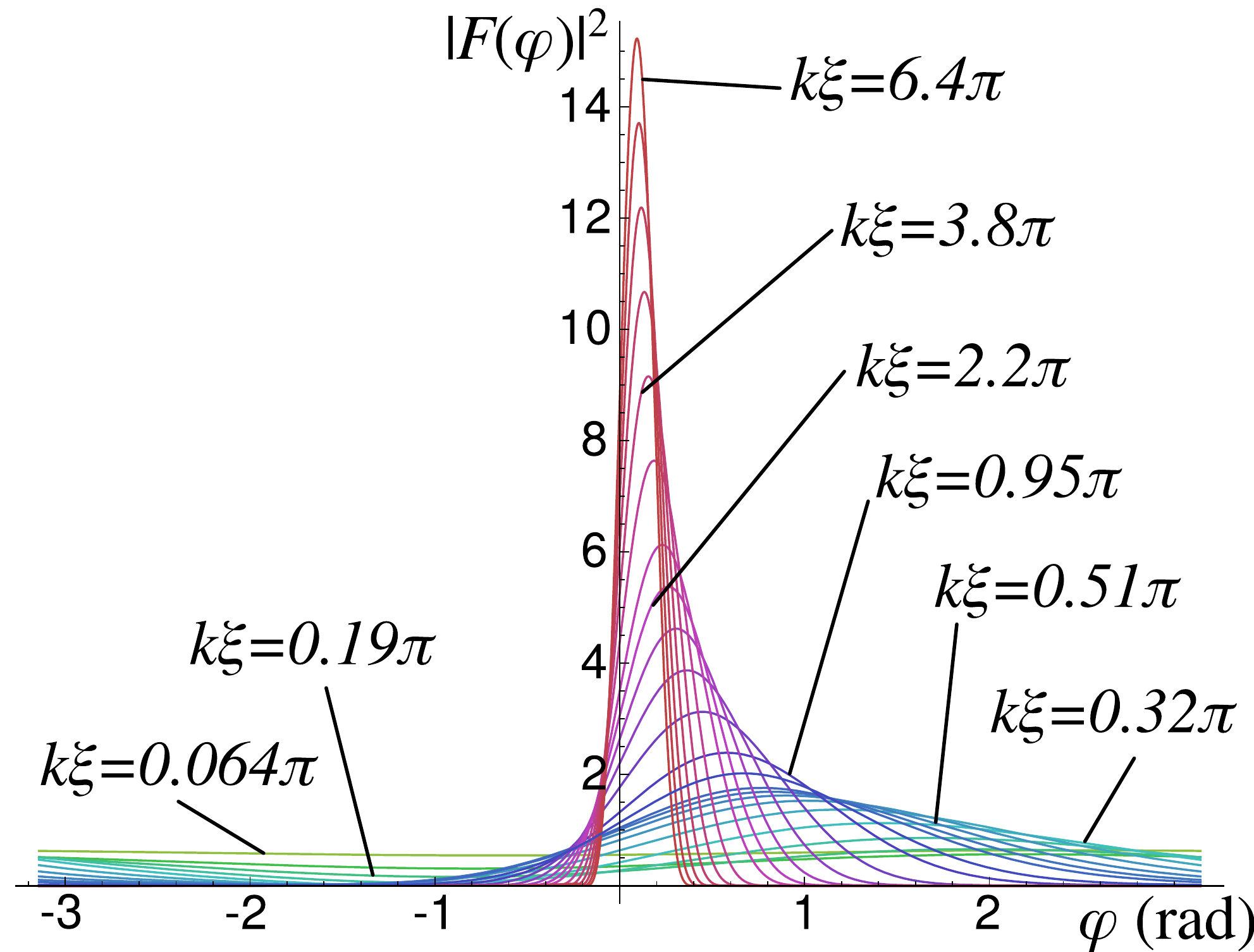}}
\end{center}
\caption{{ The scattering properties obtained by numerical and analytical calculations.}
(a) 
The Hall angles (red line) and velocities (blue line) of skyrmion motion are estimated from the numerical results for different wavenumbers $\bar{k}$. 
To obtain these values, we traced the center of mass coordinate $\bm{R}$ of a skyrmion between $Y=51$ and $Y=31$. 
The coordinate $\bm{R}$ is defined as 
$\bm{R} \equiv \int \td^2 x \rho_\mathrm{top} (\bm{r}) \bm{r} / \int \td^2 x \rho_\mathrm{top} (\bm{r})$, 
where $\rho_\mathrm{top} \equiv \bm{m}(\bm{r}) \cdot \left( \partial_x \bm{m}(\bm{r}) \times \partial_y \bm{m}(\bm{r}) \right)$.
There is a strong non-monotonic wavenumber-dependent behaviour in both quantities. We compare these observations to the idealized cases of magnons scattering off a uniform flux tube by a Aharonov--Bohm type calculation as detailed in Supplemental Information \ref{subsec:Derivation of the scattering amplitude}:
(b) Expectation of the magnon Hall angle $\phi$ as a function of wavenumber $k$. It is strongly peaked around $k\xi \approx 1$, and vanishes for both low and high wavenumber, in the latter case as $\sim 1/k$. 
(c) Magnon scattering amplitude of several wavenumbers $k$. The asymmetry in left or right scattering can be clearly seen and is due to the effective Lorentz force induced by the Berry phase of the skyrmion. For low wavenumber, the scattering amplitude is almost flat, indicating the wave `missing' or `ignoring' the skyrmion; for high wavenumber it is strongly peaked, indicating mostly forward scattering, which is well known in the Aharonov-Bohm effect.
}
\label{fig:ang}
\end{figure*}

{\em Numerical results. }
Our model is the chiral magnet on the 2D square lattice:
\begin{align}
\mathcal{H}=&-J\sum_{\bm r} {\bm m}_{\bm r} \cdot 
\left( {\bm m}_{{\bm r}+{\bm e_x}}+{\bm m}_{{\bm r}+{\bm e_y}} \right) \nonumber\\
&
-D\sum_{\bm r} \left( 
 {\bm m}_{\bm r} \times {\bm m}_{{\bm r}+{\bm e_x}} \cdot {\bm e_x}
+{\bm m}_{\bm r} \times {\bm m}_{{\bm r}+{\bm e_y}} \cdot {\bm e_y} \right) \nonumber\\
&
-B_z \sum_{\bm r} \left( {\bm m}_{\bm r} \right)_z.
\label{eqn:Hamiltonian}
\end{align}
Here, $\bm{m}_{\bm{r}}$ is the unit vector representing the direction of the local magnetic moment.
In the following, we measure all physical quantities in units of $J=\hbar=a=1$, 
where $\hbar$ is the reduced Planck constant and $a$ is the lattice constant.
We fix DM interaction $D=0.18$.
At $T=0$, the ground state is the helical state for external field $B<B_{\mathrm{c}1}=0.0075$, 
the ferromagnetic state for $B>B_{\mathrm{c}2}=0.0252$, and the skyrmion-crystal 
for $B_{\mathrm{c}1}<B<B_{\mathrm{c}2}$.

To study the scattering of magnon plane waves off a single skyrmion, we have 
performed micromagnetic simulations based on the Landau--Lifshitz--Gilbert (LLG) equation:
\begin{align}
\frac{\bm{m}_{\bm{r}}}{\td t}
=- \bm{m}_{\bm{r}} \times \bm{B}^\mathrm{eff}_{\bm{r}}
+\alpha \bm{m}_{\bm{r}} \times \frac{ \td \bm{m}_{\bm{r}}}{\td t},
\label{eqn:LLG}
\end{align}
where $\alpha$ is the Gilbert damping coefficient 
and ${\bm B}^{\rm eff}_{\bm r}=-\frac{\partial \mathcal{H}}{\partial {\bm m}_{\bm r}}$.
We perform the simulation at $B=0.0278(>B_{\mathrm{c}2})$, putting a metastable 
skyrmion at the center of ferromagnetic background (Fig.~~\ref{fig:snapshots}(a1)).
The size of the skyrmion $\xi$ in this paper is defined as the distance from the core ($m_z=-1$) 
to the perimeter ($m_z=0$), and $\xi = 8$ for our parameter set.
At the lower boundary a forced oscillation of frequency $\omega$ 
with  fixed amplitude $A \equiv \langle m_x^2+m_y^2 \rangle= 0.0669$ 
is imposed on the spins, producing spin waves with
wavevector $\bm{k} = (0,k)$ traveling toward the top.
Here, the amplitude of the magnon with wavenumber $k$ is proportional to 
$ \frac{1}{ \omega^2 - \omega_{\bf k}^2 + i \alpha \omega}$, 
where $\omega_{\bm{k}}$ is the dispersion of the magnon with energy gap $B$. 
We estimated the averaged $\bar{k}$ from the real space image of the magnon propagation. 
For $\omega = 0.08,0.04,0.02,0.0125$ and $0.01$, we find 
$\bar{k} \xi \simeq 1.87\pi,1.20\pi,0.83\pi,0.64\pi$ and $0.52\pi$, respectively. 
Note that the latter three frequencies are below the magnon gap. 

Figure~\ref{fig:snapshots} shows snapshots of the scattering processes with three 
different wavelengths (see also Supplementary Movies 1, 2 and 3).
These lead to several remarkable observations.
First, one can clearly see that the identity of the skyrmion remains intact even though some 
distortion of its shape occurs. This originates in the topological protection, and is not a trivial 
fact since both the skyrmion and magnons are made out of the same spins. Namely, the 
skyrmion number $Q = \frac{1}{4\pi} \int \td^2 x \ 
\bm{m} \cdot  \left( \partial_x \bm{m} \times \partial_y \bm{m} \right)$ is $-1$ for the 
skyrmion while that of magnons is zero, and hence the
conservation of the skyrmion number protects the identity of the skyrmion.
Second, the incident wave is clearly scattered by the skyrmion, with sizable `skew angle' or `Hall angle'. 
As the wavenumber $\bar{k}$ is increased, the diffraction becomes smaller and one can define the
trajectory of the scattered magnons clearly in Fig.~\ref{fig:snapshots} (a1)-(a4) for $\bar{k}\xi \simeq 1.87\pi$.
As shown in the blue lines in Fig.~\ref{fig:snapshots} (a4), the scattered trajectory has an angle $\bar{\varphi}$
compared with the direction of the incident magnons (vertical line).
As the wavenumber $\bar{k}$ is reduced, the diffraction is enhanced, but the skewness
of the scattered waves can still be seen in Figs.~\ref{fig:snapshots} (b2)-(b4) for $\bar{k}\xi \simeq 0.83\pi$ and (c2)-(c4) for $\bar{k}\xi \simeq 0.52\pi$. 
Therefore, the skew angle $\bar{\varphi}$ strongly depends on $\bar{k}\xi$.
Third, by tracing the center of mass position of the skyrmion, it is found that 
it moves in turn backward and sidewards in the opposite direction as indicated by the
yellow lines in Figs.~\ref{fig:snapshots} (a4),(b4), and (c4). 
The skew angle $\Phi$ of the skyrmion motion is plotted in Fig.~\ref{fig:ang} (a), which shows strong $\bar{k}$-dependence.   
Also the speed $v$ of the skyrmion depends on the wavenumber $\bar{k}$ for fixed amplitudes of
the magnons, as shown in Fig.~\ref{fig:ang} (a). 
This skyrmion motion can be interpreted as the spin transfer torque by the magnon spin current, or
equivalently analyzed from the viewpoint of momentum conservation as will be discussed below.

{\em Skyrmion momentum. }
The dynamic term of a skyrmion particle is $S_\mathrm{dyn} = \int \td t \td^2 x \ \mathcal{L}$,  where~\cite{Zang11},
\begin{align}
\mathcal{L} &=   2 \pi Q (Y \partial_t X - X \partial_t Y)
 + \frac{M}{2} \big( ( \partial_t X)^2 + (\partial_t Y)^2\big)   ,
\end{align}
where $X,Y$ are the skyrmion center of mass coordinates,
and $M$ is the mass of the skyrmion. Then the momentum is 
$P_x = \frac{\partial \mathcal{L}}{\partial \partial_t X} = 2 \pi Q Y + M \partial_t X$ and 
$P_y = \frac{\partial \mathcal{L}}{\partial \partial_t Y} = -2 \pi Q X + M \partial_t Y$. 
Assuming a  massless skyrmion and elastic scattering, we can estimate the skew angle as follows. For the magnon 
$p_\mathrm{mag}^{(\mathrm{in})} = \begin{pmatrix} 0 & k \end{pmatrix}$ and $p_\mathrm{mag}^{(\mathrm{out})} = 
\begin{pmatrix} k \sin \bar{\varphi} & 
k \cos \bar{\varphi} \end{pmatrix}$, then 
$\Delta P_\mathrm{skyrmion} = \begin{pmatrix} -k \sin \bar{\varphi} & k (1 - \cos\bar{\varphi}) \end{pmatrix}$. Using $P_x = 2 \pi Q Y, P_y = -2 \pi Q X$ 
one finds the skyrmion Hall angle $\Phi = \arctan (\Delta X/ \Delta Y) = \bar{\varphi}/2$. 
The numerics, i.e., $\Phi$ and $\bar{\varphi}$ in Figs.~\ref{fig:snapshots} (a4),
is consistent with this relation within the error bars. 
In the present simulation, the displacement $\Delta R$ of the skyrmion is about $30$, 
over the time period of $2000$ for $k\xi \simeq 1.87\pi$. 
The velocity $v$ is of the order of $30/2000 \cong 1.5 \times 10^{-2}$.
The mass $M$ is of the order of the number of spins constituting one skyrmion and is of the order of $200$ in our simulation. 
Therefore, $M v \sim 3 \ll 2 \pi \Delta R \sim 200$, and hence the assumption of the massless skyrmion above is justified. 

We can estimate the velocity of the skyrmion purely in terms of momentum transfer of the spin wave to the skyrmion. The momentum transfer per unit time is $|\Delta P_\mathrm{skyrmion}|/T_k$, where the denominator is the time it takes a magnon to pass through the skyrmion $T_k \equiv 2\xi / v_k$ where $v_k$ is the group velocity of the magnon, given by $v_k = \frac{\partial \omega_k}{\partial k} = 2Jk$, 
where $\omega_k = Jk^2 + B$ is the magnon dispersion. With the parameters used in our simulations, we find for instance for the case of $k\xi = 1.87\pi$ that $v = 0.0058$ which is comparable to the value we obtain in the simulations $v = 0.015$ in Fig.~\ref{fig:ang}(a). (For details see Supplemental Information \ref{subsec:Skyrmion velocity by momentum transfer}.) These simple momentum conservation considerations lead us to conclude that the skyrmion 
is a particle with well-defined momentum, that nevertheless defies the Newtonian intuition. 
For instance, here an elastic scattering process causes backwards motion of the skyrmion, 
which is impossible for Newtonian particles.

{\em  Effective magnetic field.}
To further identify the nature of the magnon skew scattering, we map the situation onto that of a charged particle (the magnon) moving in the background of a static magnetic field (the skyrmion), assuming the disturbances of the magnon on the emergent fictitious magnetic field are small. The emergent field corresponds to the skyrmion number, so the sign of the scattering direction is fixed, but would be opposite for an antiskyrmion configuration. This corresponds precisely to Aharonov--Bohm (AB) scattering, and using results from the extensive literature, we derive an exact expression for the scattering amplitude of the magnon (see Supplemental Information \ref{subsec:Derivation of the scattering amplitude}),
\begin{equation}\label{eq:scattering amplitude}
 F(\varphi) = f^\mathrm{AB}(\varphi) + \frac{\te^{-\ti \pi / 4}}{\sqrt{2\pi k}} \sum_{n=-\infty}^{\infty} \te^{ 
\ti \pi (n - |n
+ Q|)} \big( \te^{2\ti \Delta_n} - 1\big) \te^{\ti n \varphi}.
\end{equation}
Here the AB contribution $f^\mathrm{AB}(\varphi)$ vanishes for integer skyrmion number $Q$, and $\Delta_n$ is the phase shift of the $n$th partial wave. 
The scattering amplitude is evaluated numerically; the results are shown in Fig.~\ref{fig:ang} (b) and (c). 
We find a large skew scattering that is strongly wavenumber-dependent, up to 60$^\circ$ around $k\xi=1$. This is consistent with the numerical results, and demonstrates that the skew-scattering is due to the emergent magnetic field or Berry phase of the skyrmion.

{\em Conclusions. }
We have studied the scattering process of magnons and a skyrmion both numerically and
analytically. We have found a large skew angle of the scattering due to the emergent effective magnetic field, and the skyrmion motion 
as the back action of the scattering. This process strongly depends on the wavenumber $k$ of
the magnons times the size $\xi$ of the skyrmion, and the skew angle can be 
as large as 60 degree when $k \xi \cong 1$. It can be viewed as an elastic scattering process when taking into account the peculiar momentum of the skyrmion motion. 
This should be compared with the case of topological Hall effect
of the conduction electrons coupled to the skyrmions~\cite{Lee09,Neubauer09,Kanazawa11,vanHoogdalem13},
where the Hall angle is typically of the order of $10^{-3}$ because 
the fermi wavenumber $k_F$ of the electrons is much larger than $\xi^{-1}$.

{\em Acknowledgments. }
This work was supported by Grant-in-Aids for Scientific Research (Nos.~24224009, 25870169, 25287088) 
from the Ministry of Education, Culture, Sports, Science and Technology (MEXT) of Japan, 
Strategic International Cooperative Program (Joint Research Type) from Japan Science and Technology Agency, 
and by Funding Program for World-Leading Innovative R\&D on Science and Technology (FIRST Program).  
A.J.B. is supported by the Foreign Postdoctoral Researcher program at RIKEN.


\section*{SUPPLEMENTAL MATERIAL}
\subsection{Skyrmion velocity by momentum transfer}\label{subsec:Skyrmion velocity 
by momentum transfer}
Here we derive an estimate for the velocity of the skyrmion by transfer of momentum 
from the magnons to the skyrmion.

A plane wave $\sqrt{A} \te^{-\ti \omega t + \ti \bar{k}y}$ has momentum 
$p^{(\mathrm{in})} = A \bar{k}$. The incoming magnons of average wavenumber $\bar{k}$ are generated by a forced oscillation with magnitude $A \equiv \langle m_x^2+m_y^2 \rangle= 0.0669$ per lattice spin. The part of the incident wave that interacts with the skyrmion is of size $2\xi$, the diameter of the skyrmion.  Hence the momentum of the part of the magnon plane wave interacting with the skyrmion is $k = 2\xi A \bar{k}$.

By conservation of momentum, the momentum transfer of the spin wave with momentum $\mathbf{p}_\mathrm{mag}^{(\mathrm{in})} = \begin{pmatrix} 0 & k \end{pmatrix}$ and $\mathbf{p}_\mathrm{mag}^{(\mathrm{out})} = 
\begin{pmatrix} k \sin \bar{\varphi} & k \cos \bar{\varphi} \end{pmatrix}$ is
$\Delta \mathbf{P}_\mathrm{skyrmion} = \begin{pmatrix} -k \sin \bar{\varphi} & k (1 - \cos\bar{\varphi}) \end{pmatrix}$. The magnitude of the skyrmion momentum is $|\Delta \mathbf{P}_\mathrm{skyrmion}| = k \sqrt{2 - 2 \cos\bar{\varphi}}  = 2 k \sin \bar{\varphi}/2= 4\xi A \bar{k} \sin \bar{\varphi}/2$.

Now we are sending in a continuous plane wave instead of a single magnon. The time it takes 
for the plane wave to pass by/through the skyrmion is $T_k \equiv 2\xi / v_k$ where $v_k$ 
is the group velocity of the magnon, given by $v_k = \frac{\partial \omega_k}{\partial k} = 2Jk$, 
where $\omega_k = Jk^2 + B$ is the magnon dispersion. Hence in one unit of time, the plane 
wave interacts with $1/T_k$ part of the skyrmion. Thus the amount of momentum transferred 
in one unit of time is
\begin{equation}
\Delta \tilde{P} \equiv  \frac{ |\Delta \mathbf{P}_\mathrm{skyrmion}| }{T_k} = 
\frac{ 4\xi A \bar{k} \sin \bar{\varphi}/2 }{ 2\xi / 2J \bar{k} } =    4J A \bar{k}^2 \sin \bar{\varphi}/2 .
\end{equation}

In our units $J = 1$. The forced oscillation in our simulations has a magnitude $A =  0.0669$. 
For the case of $\bar{k}\xi  = 1.87\pi$ ($\bar{k} = 1.87\pi/\xi = 1.87\pi/8 \approx 0.73$) we find 
$\phi/2 \approx 15^\circ$, so $\sin \phi/2 \approx 0.26$ (see Fig.~\ref{fig:snapshots}(a4)). 
In this case we therefore find  $\Delta \tilde{P} \approx 0.036$ and skyrmion velocity 
$V = \Delta \tilde{P} / 2\pi = 0.0058$. This is different from the
value obtained in the simulations ( 0.015 ) by a factor of $\cong 2.5$ (Fig. \ref{fig:ang}a), but considering 
the rough and tentative nature of the estimate, the agreement is rather good.

\subsection{Derivation of the scattering amplitude}\label{subsec:Derivation of the scattering amplitude}
Here we derive Eq. \eqref{eq:scattering amplitude}. We will map the magnon--skyrmion scattering onto a charged particle moving in the background of a static magnetic field, and use results from the Aharonov--Bohm (AB) effect to obtain an exact expression for the scattering amplitude of the magnon.

In the continuum limit, the Hamiltonian Eq. \eqref{eqn:Hamiltonian} for the local moments 
$\mathbf{m}(x,y)$ reads
\begin{equation}
 \mathcal{H} = \int \td^2 x \ \big[ 
 J (\nabla \mathbf{m})^2 
 + D \mathbf{m} \cdot (\nabla \times \mathbf{m}) 
 - \mathbf{B} \cdot \mathbf{m}
 \big].
\end{equation}
We can make a change of variables to a complex 2-vector 
$z_\rho = \begin{pmatrix} z_\ua & z_\da \end{pmatrix}$ (a $CP(1)$-field) 
via $\mathbf{m} = z^*_\rho \bm{\sigma}_{\rho\sigma} z_\sigma$, 
where $\bm{\sigma}_{\rho\sigma}$ are the Pauli matrices and the 
constraint $\sum_\rho |z_\rho|^2 = 1$ must be imposed. The Hamiltonian turns into
\begin{equation}
 \mathcal{H} = \int \td^2 x \ 
 4J | ( \nabla + \ti \mathbf{a} + \ti \kappa \bm{\sigma} ) z_\rho |^2 
 - \mathbf{B} \cdot z^*_\rho \bm{\sigma}_{\rho\sigma} z_\sigma,
\end{equation}
where $\kappa = D/4J$ and $\mathbf{a} = \ti z^*_\rho \nabla z_\rho$. 
The Hamiltonian is invariant under gauge transformations 
$z_\rho \to z_\rho \te^{\ti \varepsilon}$ and 
$\mathbf{a} \to \mathbf{a} + \nabla \varepsilon$, where $\varepsilon(\mathbf{r})$ 
is any smooth scalar field.
The gauge field is related to the Berry curvature $\mathbf{b} = \nabla \times \mathbf{a}$, 
and the skyrmion number $\alpha = \frac{1}{4\pi} \int \td^2 x \ \mathbf{b} = \frac{1}{4\pi} \int \td^2 x \ 
\bm{m} \cdot  \left( \partial_x \bm{m} \times \partial_y \bm{m} \right) \equiv Q$ 
is quantized. We now separate $z_\rho = \breve{z}_\rho + z^0_\rho$ into magnon and 
skyrmion contributions, and assume that a static skyrmion $\mathbf{a}^0$ of size $\xi$ with 
$\alpha = -1$ has formed while the magnons $\breve{z}_\rho$ move in this skyrmion 
background. A typical skyrmion solution in polar coordinates is 
$a_r=0,\quad a_\varphi = \frac{r}{\xi^2+r^2}$. For small deviations from this background 
configuration we need only to consider the exchange term; the DM and Zeeman contributions 
are constant on this energy scale. Summarizing, we are considering the low-energy dynamics of
\begin{equation}\label{eq:low-energy Hamiltonian}
 \mathcal{H}_\mathrm{LE} = \int \td^2 x \ 
 4J | ( \nabla + \ti \mathbf{a}^0) \breve{z}_\rho |^2.
\end{equation}
This is precisely the Hamiltonian of a charged particle moving in an external magnetic field 
$\mathbf{b}^0 = \nabla \times \mathbf{a}^0$. Notice that the components 
$\breve{z}_\ua$, $\breve{z}_\da$ are now decoupled at this level of 
the approximation. We are interested in the scattering 
outcome of an incoming plane wave, far away from the origin of the skyrmion. 
Then in this ferromagnetic regime, the spins point along the out-of-plane $z$-direction, 
and we can make the approximation $z_\ua \approx 1$. In other words, we only consider the 
field $\breve{z}_\da$. The problem of a charged particle scattered by a magnetic flux was intensively 
studied in and after the discovery of the Aharonov--Bohm (AB) effect 
\cite{Aharonov59,Kretzschmar65,Olariu85,Brown85,Brown87}. 
Using earlier results \cite{Kretzschmar65,Olariu85,Brown85,Brown87},
we shall approximate our smooth skyrmion potential $a_\varphi = r/(\xi^2+r^2)$ with that of a 
uniform magnetic flux:
\begin{equation}\label{eq:flux tube vector potential}
 a_\varphi = \begin{cases} 1/ r, &\quad r \ge \xi \\ r / \xi^2, &\quad r \le \xi \end{cases} .
\end{equation}
One can verify that the total fictitious flux $\alpha =\int \nabla \times \mathbf{a}$ is the same 
for both potentials (in the AB-setup, the value of $\alpha$ corresponds to the product of the 
electric charge and magnetic flux). As the magnon will principally scatter due to the fictitious 
Lorentz force, this approximation will not deviate too much from the actual situation, and has 
the advantage of allowing for an exact solution. 

 Now we derive the exact solution of Eq. \eqref{eq:low-energy Hamiltonian} for a tube of uniform magnetic 
flux Eq. \eqref{eq:flux tube vector potential}, following Brown \cite{Brown85,Brown87}. In AB scattering one is usually interested in the case that the 
particle does not enter regions of finite magnetic flux, but nevertheless the case of a uniform magnetic 
flux tube of radius $\xi$ has 
been analyzed  in Refs. \cite{Kretzschmar65,Olariu85,Brown85,Brown87}. They consider an electrostatic 
shielding potential 
$V$ to prevent the particle from entering the region of non-zero flux, but the results are in 
fact general for any $V$, and the limit of $V \to 0$ may be taken without additional treatment, 
as we do from now on.

For $\mathbf{a}^0 = 0$ Eq. \eqref{eq:low-energy Hamiltonian} describes plane waves of energy $E = a^2 J k^2$, 
where $a$ is the lattice constant and $k$ is the wavenumber. For non-zero $\mathbf{a}^0$ the equation 
of motion for a particle of this energy reads in polar coordinates,
 \begin{equation}
 \left[ \partial_r^2 + \frac{1}{r} \partial_r + \frac{1}{r^2} \big( \partial_\varphi - \ti r (-\alpha)  a_\varphi\big)^2  
+ k^2 \right] \breve{z}_\da = 0.
\end{equation}
Here we tentatively allow the skyrmion number $\alpha$ to deviate from the value $-1$. The only term dependent on 
$\varphi$ is the one involving $\partial_\varphi$, and we can make a partial wave expansion 
$\breve{z}_\da (r,\varphi) = \sum_n \breve{z}_n =  \sum_n w_n(r) \te^{\ti n \varphi}$. For $r \ge \xi$, the $w_n$ 
are eigenfunctions of the equation,
\begin{equation}
 \left[ \partial_r^2 + \frac{1}{r} \partial_r + k^2  \frac{1}{r^2} \left( n  +  \alpha \right)^2   \right] w_n^{>}  = 0.
\end{equation}
This is precisely Bessel's equation, and the general solution is
\begin{equation}\label{eq:general outside solution}
 \breve{z}^{>}_n = \te^{\ti n \varphi} \left[ a_n J_{|n+\alpha|} (kr) + b_n Y_{|n+\alpha|}(kr) \right].
\end{equation}
For the region $r \le \xi$, the Schr\"odinger equation reads,
\begin{equation}
 \left[ \partial_r^2 + \frac{1}{r} \partial_r - \frac{1}{r^2} \left( n +   
\alpha \frac{r^2}{\xi^2} \right)^2  + k^2 \right] w^<_n(r) = 0.
\end{equation}
We make a change of variables $v = \alpha r^2 / \xi^2$ and $f_n(v) = r w^<_n(r)$. 
The above equation is then rewritten as,
\begin{equation}
 \left[ \partial_v^2 + \frac{1/4 - n^2/4}{v^2} + \frac{k^2 \xi^2 / 4 \alpha - n/2}{v} - \frac{1}{4} \right] f_n(v) = 0.
\end{equation}
This is known as Whittaker's equation for the parameters $\kappa = k^2 \xi^2 / 4 \alpha - n/2$ 
and $\mu^2 =  n^2/4$. 
The solutions, known as Whittaker functions $M_{\kappa,\mu}(v)$, are not well defined for $\mu = -1,-2,\ldots$, 
but for our purposes it suffices to choose $\mu = |n/2|$. These solutions are
\begin{equation}\label{eq:Whittaker function}
 f_n(v) = M_{\kappa,\mu} (v) = \te^{-z/2} z^{\mu + 1/2} \Phi(\tfrac{1}{2} + \mu - \kappa,2\mu +1, v),
\end{equation}
where $\Phi$ is the confluent hypergeometric series,
\begin{equation}
 \Phi(a,c,v) = 1 + \frac{a}{c} v + \frac{a(a+1)}{c(c+1)} \frac{1}{2!} v^2 + \ldots.
\end{equation}
Continuity in the wavefunction and its first derivative at the matching point $r = \xi$ leads to the equalities,
\begin{align}
 c_n w^<_n (\xi) &= a_n J_{|n+\alpha|} (k\xi) + b_n Y_{|n+\alpha|} (k\xi), \label{eq:boundary continuity}\\
 \Big[ c_n \partial_r w^<_n (r) &= a_n \partial_r J_{|n+\alpha|} (kr) +  b_n 
\partial_r Y_{|n+\alpha|} (kr)\Big]_{r=\xi}.\label{eq:boundary derivative continuity}
\end{align}
With the notation $\Phi_{\kappa,\mu}(v) = \Phi(\tfrac{1}{2} + \mu - \kappa,2\mu +1, v)$ one can derive,
\begin{equation}
 \left. \partial_r w^<_n \right|_{r=\xi} = \frac{ M_{\kappa,\mu}(\alpha)}{\xi^2} \left( |n| -\alpha + 
2\alpha \frac{\left.\partial_v \Phi_{\kappa,\mu}(v)\right|_{v=\alpha}}{\Phi_{\kappa,\mu}(\alpha)}  \right).
\end{equation}
Substituting Eq. \eqref{eq:boundary continuity} in Eq. \eqref{eq:boundary derivative continuity} we eventually find,
\begin{equation}\label{eq:b div a}
 \frac{b_n}{a_n} = - \frac{A_n J_{|n+\alpha|} - \left.\partial_{\bar{r}} J_{|n+\alpha|}(\bar{r})\right|_{\bar{r} 
= k\xi}}{A_n Y_{|n+\alpha|} -  \left.\partial_{\bar{r}} Y_{|n+\alpha|}(\bar{r})\right|_{\bar{r} = k\xi}},
\end{equation}
where we have defined,
\begin{equation}\label{eq:An coefficients}
 A_n = \frac{1}{k\xi} \left( |n| -\alpha + 2\alpha \frac{\left.\partial_v 
\Phi_{\kappa,\mu}(v)\right|_{v=\alpha}}{\Phi_{\kappa,\mu}(\alpha)}  \right).
\end{equation}
We expect to retrieve the Aharonov--Bohm result,
\begin{equation}
 \breve{z}^{\mathrm{AB}} =  \sum_{n=-\infty}^{\infty} \te^{\ti n \varphi} \te^{\ti \delta_n^\mathrm{AB}} J_{|n+\alpha|}(kr).
\end{equation}
where $\delta_\mathrm{AB} = - |n+\alpha|\pi/2$, in the limits of vanishing skyrmion 
size $\xi \to 0$ or vanishing flux $\alpha \to 0$. Brown \cite{Brown87} has shown that the solution
\begin{align}\label{eq:partial wave coefficients}
 a_n &= \cos \Delta_n \te^{\ti \Delta_n} \te^{\ti \delta_\mathrm{AB}} & b_n &
= \sin \Delta_n \te^{\ti \Delta_n} \te^{\ti \delta_\mathrm{AB}},
\end{align}
corresponds to an incoming plane wave and an outgoing propagating scattered wave, and this solution 
does reduce to the AB results in the mentioned limits, for which all $\Delta_n \equiv \tan (-b_n/a_n) \to 0$. 
Brown has also shown that, for any $\alpha$, $\Delta_n \to 0$ as $n \to \infty$, and in practice the $\Delta_n$ 
vanish quickly for $n > k\xi$. Writing the solution as the superposition of an incoming and a scattered wave, 
$\breve{z}^> = \exp(\ti kx) + F(\varphi)\frac{\exp(ikr)}{\sqrt{r}}$, Brown obtains the scattering amplitude,
\begin{equation}\label{eq:scattering amplitude SI}
 F(\varphi) = f^\mathrm{AB}(\varphi) + \frac{\te^{-\ti \pi / 4}}{\sqrt{2\pi k}} \sum_{n=-\infty}^{\infty} \te^{ 
\ti \pi (n - |n
+\alpha|)} \big( \te^{2\ti \Delta_n} - 1\big) \te^{\ti n \varphi}.
\end{equation}
Here $f^\mathrm{AB}$ is the Aharonov--Bohm scattering amplitude,
\begin{equation}\label{eq:AB scattering amplitude}
 f^\mathrm{AB}(\varphi)  \frac{\te^{\ti \pi /4}}{\sqrt{2\pi k}} \sin ( \pi |\alpha|) \frac{\te^{\ti \varphi  
\mathrm{sgn}(\alpha)} }{\cos(\tfrac{1}{2} \varphi)}.
\end{equation}
The AB scattering amplitude is clearly vanishing for integer $\alpha$.


We evaluate this exact solution Eq. \eqref{eq:scattering amplitude SI} numerically. 
Here we make use of the fact that the phase shifts $\Delta_n$ tend to zero quickly for $n > k \xi$, 
meaning that only the lowest few partial waves contribute to scattering. The scattering amplitude 
and the skew angle $\bar{\varphi} = \int \varphi |F(\varphi)|^2 / \int |F(\varphi)|^2$ for several 
values of $\bar{k}\xi$ are shown in Figs. \ref{fig:ang}. We clearly see a large 
skew angle at the scattering of the magnon for $k \approx 1/\xi$. For both very low and very high wavenumber 
the skew angle tends to zero, and the maximum skew angle is about $60^\circ$ around $\bar{k}\xi \approx 1.1$. This should be compared with the case of topological Hall effect
of the conduction electrons coupled to the skyrmions~\cite{Lee09si,Neubauer09si,Kanazawa11si},
where the Hall angle is typically of the order of $10^{-3}$ because 
the Fermi wavenumber $k_F$ of the electrons is much larger than $\xi^{-1}$. 
 For larger $k$, the skew angle is reduced and asymptotically 
behaves as $\propto 1/k$. This indicates that the velocity of the skyrmion induced by the
back action should be saturated in Fig.~\ref{fig:ang} since the 
momentum transfer from magnons to skyrmion is $\propto k \times \Phi \sim k \times 1/k \sim $ constant 
for large $k$ limit assuming the elastic scattering. Unfortunately, this large $k$ region was not successfully 
analyzed in the numerical simulation due to a technical difficulty, which requires further studies.

\end{document}